\documentclass[%
 reprint,
superscriptaddress,
 amsmath,amssymb,
 aps,
]{revtex4-2}

\usepackage{graphicx}
\usepackage{dcolumn}
\usepackage{bm}

\usepackage[utf8]{inputenc}
\usepackage[T1]{fontenc}
\usepackage{mathptmx}
\usepackage{etoolbox}
\usepackage[svgnames]{xcolor,pict2e}
\usepackage[skins]{tcolorbox}

\makeatother
\begin{document}

\preprint{APS/123-QIS}

\title{High-Efficiency, High-Fidelity Charge Initialization of Shallow Nitrogen Vacancy Centers in Diamond}
\author{Marjana Mahdia}
\affiliation{Department of Electrical and Computer Engineering, Princeton University, Princeton, New Jersey 08544, USA}
\author{Artur Lozovoi}
\affiliation{Department of Electrical and Computer Engineering, Princeton University, Princeton, New Jersey 08544, USA}
\author{Jared Rovny}%
\affiliation{Department of Electrical and Computer Engineering, Princeton University, Princeton, New Jersey 08544, USA}
\author{Zhiyang Yuan}%
\affiliation{Department of Electrical and Computer Engineering, Princeton University, Princeton, New Jersey 08544, USA}
\author{Carlos A. Meriles}%
\affiliation{Department of Physics, CUNY-City College of New York, 160 Convent Avenue New York, New York 10031, USA}
\author{Nathalie P. de Leon}%
 \altaffiliation[]{Contact author: npdeleon@princeton.edu}
\affiliation{Department of Electrical and Computer Engineering, Princeton University, Princeton, New Jersey 08544, USA}

\date{\today}

\begin{abstract}

Nitrogen vacancy (NV) centers in diamond exhibit long spin coherence times, optical initialization, and optical spin readout under ambient conditions, making them excellent quantum sensors. However, the conventional scheme for charge state initialization based on off-resonant green excitation results in significant state preparation errors, typically around 30$\%$. One method for improving charge state initialization fidelity is to use multicolor excitation, which has been demonstrated to achieve a near-unity preparation fidelity for bulk NV centers by using a few milliseconds of near-infrared (5 mW) and green (10 $\mu$W) excitation. The translation of such schemes to NV centers near the diamond surface with higher efficiency optical pumping would enable myriad tasks in nanoscale sensing. Here, we demonstrate a protocol for efficient charge initialization of shallow NV centers between 5 nm and 15 nm from the diamond surface. By carefully studying the charge dynamics of shallow NV centers, we identify a region of parameter space that allows for near-unity (95\%) charge initialization within 300 $\mu$s of near-infrared (1 mW) and green (10 $\mu$W) excitation. The time to 90\% charge initialization can be as fast as 10 $\mu$s for 4 mW of near-infrared and 39 $\mu$W of green illumination. This fast, efficient charge initialization protocol will enable nanoscale sensing applications where state preparation errors currently prohibit scaling, such as measuring higher-order multi-point correlators.

\end{abstract}
\maketitle

\section*{I. Introduction}

NV centers are excellent quantum sensors because of their long spin coherence times and optical spin interface at room temperature \cite{childress2013diamond,doherty2013nitrogen,schirhagl2014nitrogen}. Under photoexcitation, NV centers dynamically interconvert between their negative (NV$^-$) and neutral (NV$^0$) charge states \cite{aslam2013photo,mahdia2023probing,wood2024wavelength}; the negative charge state is the one that is used for quantum information processing and sensing. Specifically, under off-resonant green excitation (510-540 nm), which is typically used for spin readout, the steady state negative charge population is 70\% \cite{aslam2013photo}, representing the dominant source of state preparation and measurement (SPAM) error. This SPAM error reflects the detailed balance of photoionization and recombination under a particular optical excitation scheme. Using a different excitation protocol can potentially offer a way to reduce error by favoring the preparation of NV$^-$. For instance, the combination of a short (150 ns) high-power (21 mW) green illumination pulse followed by a longer (90 $\mu$s) low-power (6 $\mu$W) green pulse has been shown to increase both charge and spin initialization fidelity of the NV center \cite{wirtitsch2023exploiting}. In another example \cite{hopper2016near}, multicolor excitation consisting of simultaneous green and high-power (5 mW) near-infrared (NIR) illumination of several milliseconds has been shown to improve bulk NV$^-$ preparation fidelity. However, both methods require relatively high optical power or prolonged illumination, limiting their utility for sensing applications. Furthermore, the translation of such schemes to shallow NV centers (within 20 nm of the diamond surface) can be complicated by deleterious charge dynamics driven by surface trap states \cite{dhomkar2016long,stacey2019evidence,peace}.\newline
\indent Here we demonstrate that combined NIR and green excitation can efficiently prepare the negative charge state of shallow NV centers, between 5 nm and 15 nm from the diamond surface. Through careful study and calibration of initialization laser power and duration, we achieve near-unity (95\%) NV$^-$ initialization at short illumination time and low optical power down to 300 $\mu$s with 0.75 mW of NIR excitation. The initialization time can be further reduced, down to 10 $\mu$s, and we achieve 90\% NV$^-$ initialization fidelity at the expense of a slightly higher NIR power of 4 mW. This increased charge state preparation fidelity translates to a commensurate increase in optical spin contrast and therefore increased magnetometer sensitivity. This charge preparation scheme will be particularly advantageous for applications where scaling is hampered by SPAM errors. We demonstrate this by calculating the expected improvement for multi-point covariance magnetometry \cite{rovny2022nanoscale}.

\section*{II. Results and Discussion}
\subsection*{A. Charge state characterization}

\begin{figure*}
\centering
\includegraphics{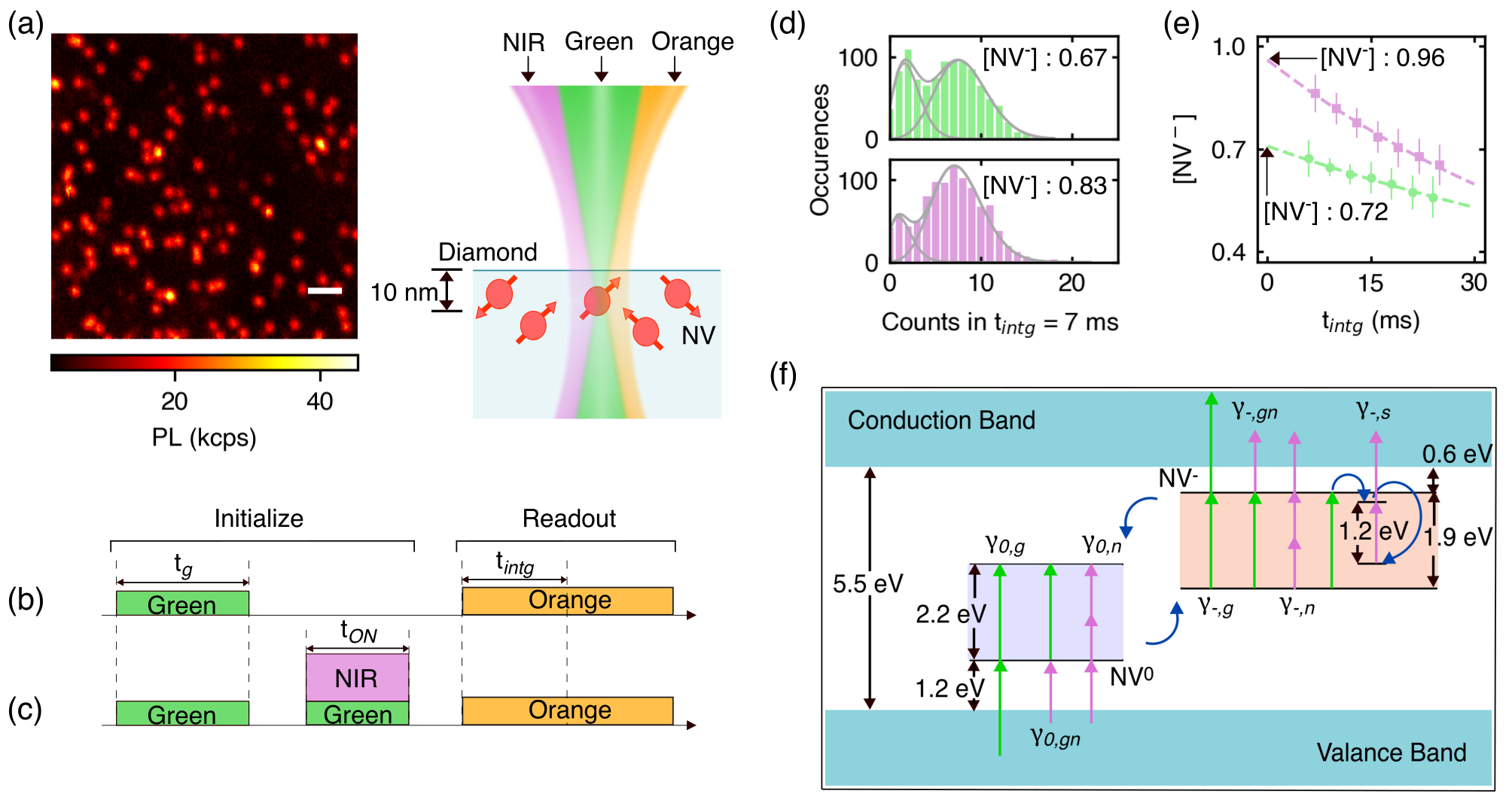}
\caption{\textbf{Multicolor excitation of NV centers.} (a) Photoluminescence (PL) confocal image of NV centers (left). The white scale bar is 1 $\mu$m. Schematic (right) showing the experimental setting where shallow NV centers are optically excited with green (520 nm), NIR (905 nm), and orange (594 nm) laser illumination. (b), (c) Pulse sequences showing NV center charge initialization with conventional green and multicolor (simultaneous green and NIR) illumination followed by the charge state readout under low-power orange excitation. t$_g$ indicates the green laser initialization time, t$_{ON}$ indicates the simultaneous green and NIR illumination time, and t$_{\text{intg}}$ indicates the integration time window. (d) Histogram of photon counts for green (upper) and multicolor (lower) initialization. Inset text shows the population in the negative state, [NV$^-$], calculated from a double-Poisson fit. (e) [NV$^-$] as a function of t$_{intg}$, for green (green circles) and multicolor (purple squares) initialization. Extrapolated [NV$^-$] at t$_{intg}$ = 0 from an exponential fit (dashed lines) are indicated. The parameters for this measurement are t$_g$ = 4 ms, t$_{ON}$ = 10 $\mu$s, g = 32 $\mu$W, and n = 7.4 mW. A long t$_g$ is applied to ensure the reset of the population after the destructive charge readout. (f) NV center energy level diagram showing the possible transitions induced by green (green arrows) and NIR (purple arrows) illumination. The blue single-ended arrows represent inter-conversion between the NV center charge states and transition within the NV$^-$ singlet manifold. $\gamma$ represent rate coefficients for the ionization and the recombination processes.}
\label{fig:figure1}
\end{figure*}

First we examine the effect of simultaneous green and NIR illumination on near-surface NV centers. We perform nitrogen ion implantation on an electronic grade diamond and use a previously established \cite{peace} surface treatment protocol that results in favorable charge and spin properties under oxygen termination for NV centers that are between 5 nm and 15 nm from the surface (see Supplementary Material \cite{noauthor_see_nodate} for more details on sample preparation). A confocal microscope scan under green excitation (Fig. \ref{fig:figure1}(a)) shows photoluminescence from single NV centers in the sample. We measure the charge state of a typical shallow NV center using weak orange (594 nm) illumination after green (520 nm) initialization (Fig. \ref{fig:figure1}(b)) and after a simultaneous green and NIR (905 nm) multicolor initialization (Fig. \ref{fig:figure1}(c)). Histograms of the photon counts over a 7 ms detection window (Fig. \ref{fig:figure1}(d)) show two peaks corresponding to the two NV center charge states; NV$^0$ is only weakly excited by orange illumination and gives a lower average count rate, while the bright state corresponds to NV$^-$. Fitting the histogram to two Poissonian distributions allows us to quantify the population in the negative state, [NV$^-$], which we measure to be 67\% and 83\% after conventional and the multicolor initialization, respectively (see Supplementary Material \cite{noauthor_see_nodate} for more details on charge state properties after conventional and multicolor excitations).\\
\indent
The populations extracted from the histograms do not reflect the true state preparation fidelity because the orange illumination used for charge state readout is destructive. This is evident in Fig. \ref{fig:figure1}(e), where [NV$^-$] is plotted as a function of varying integration time, t$_{intg}$, for the same experiments shown in Fig. \ref{fig:figure1}(b, c). The decay of [NV$^-$] after both initialization protocols is caused by a two-photon ionization of NV$^-$ during the readout \cite{aslam2013photo}. We therefore fit the decay to an exponential function and extrapolate the data to t$_{intg}$ = 0 in order to measure the true NV$^-$ initialization fidelity: 72$\pm$2\% for green initialization and 96$\pm$2\% for multicolor initialization. We report the extrapolated [NV$^-$] in the text and the figures from here onward. The increase in NV$^-$ preparation fidelity after multicolor illumination arises from a new steady state solution to a combination of multiple ionization and recombination processes (Fig. \ref{fig:figure1}(f)): ionization with two green photons ($\gamma_{-,g}$), one green and one NIR photons ($\gamma_{-,gn}$), three NIR photons ($\gamma_{-,n}$), and one green and two NIR photons from NV$^-$ singlet manifold ($\gamma_{-,s}$); and recombination with two green photons ($\gamma_{0,g}$), one green and one NIR photons ($\gamma_{0,gn}$), and three NIR photons ($\gamma_{0,n}$).\\
\indent
\textit{Hopper et al.} \cite{hopper2016near} reported a similar NV$^-$ initialization fidelity between 91\% and 95\%, when accounting for the destructive readout, using a simultaneous green (532 nm) and broadband NIR (900-1000 nm) illumination of bulk NV centers ($\sim$3 $\mu$m deep) in a solid-immersion lens (SIL). Here, we use shallow NV centers (5 - 15 nm) in a bulk diamond substrate and achieve significantly faster rates (10 $\mu$s vs several ms) using comparable NIR power (7.4 mW vs 5 mW).

\begin{figure*}
\centering
\includegraphics{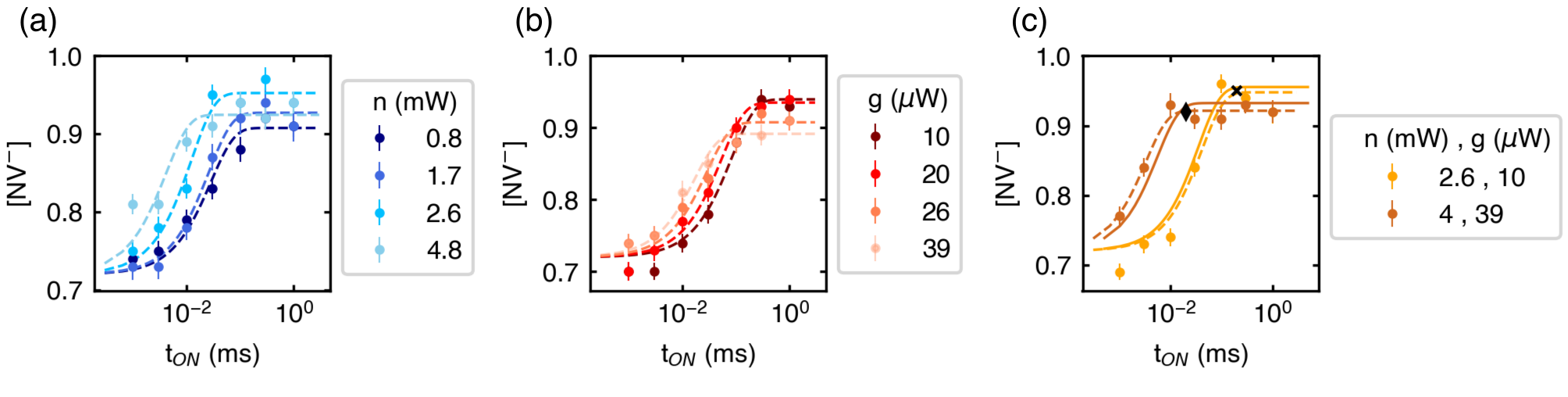}
\caption{\textbf{Parametric analysis and modeling.} (a) [NV$^-$] as a function of t$_{ON}$ for 4 different NIR powers at fixed green power g = 24 $\mu$W and (b) for 4 different green powers at fixed NIR power n = 0.75 mW. (c) [NV$^-$] as a function of t$_{ON}$ for two combinations of the NIR and the green power with the steady-state [NV$^-$] reaching 0.95 at t$_{ON}$ = 181 $\mu$s, and 0.92 at t$_{ON}$ = 17 $\mu$s, respectively. Dashed lines are fits to the data with $\rho_0$ = 0.72, and solid lines are from the model incorporating measured rates from many NV centers. The black cross and diamond markers refer to the particular sets of parameters labeled similarly in Fig. \ref{fig:figure3}(e).}
\label{fig:figure2}
\end{figure*}

\subsection*{B. Parametric analysis of charge state interconversion rates}
\begin{figure}
\centering
\includegraphics{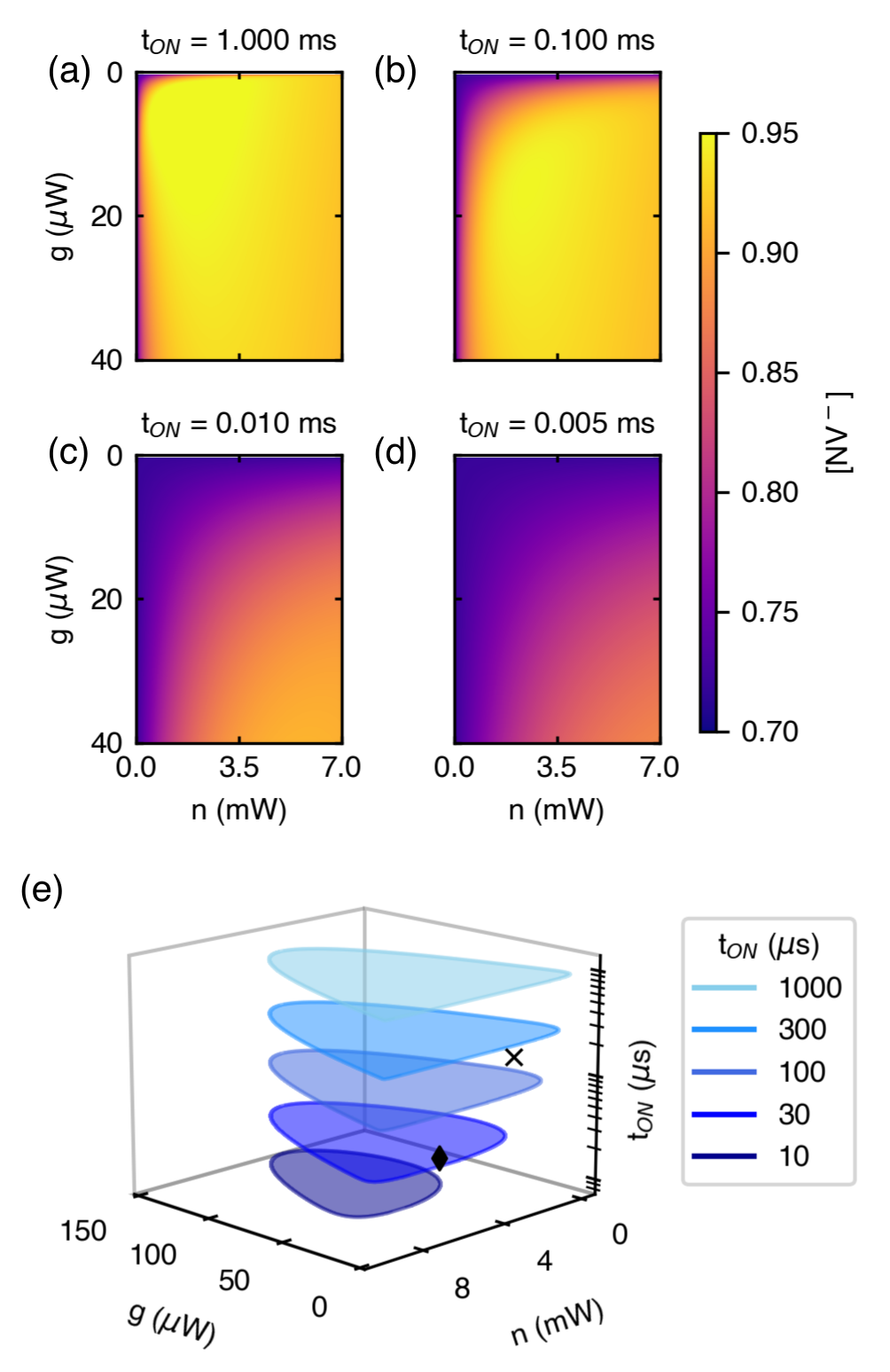}
\caption{\textbf{2D and 3D parameter space.} (a-d) 2D colormaps of [NV$^-$] as a function of the green and the NIR power for 4 values of the initialization time. (e) Stacked 2D plots outlining the parameter space volume where [NV$^-$] $\geq$ 0.90 for a set of t$_{ON}$ values. The black pointers (cross and diamond) at 200 and 20 $\mu$s correspond to the parameters and the experimental data in Fig. \ref{fig:figure2}(c).}
\label{fig:figure3}
\end{figure}

The shorter initialization time we achieve with comparable excitation power motivates a detailed analysis of the NV center photoexcitation dynamics in order to find a protocol for efficient charge state preparation. To model the dynamics of the NV center under multicolor excitation, we solve the differential equation that describes the cycling of the NV charge \cite{hopper2016near}:
\begin{equation}
\frac{d}{dt}\begin{bmatrix}
[NV^-](t)\\
[NV^0](t)
\end{bmatrix}=\begin{bmatrix}
-\gamma_i & \gamma_r\\
\gamma_i & -\gamma_r
\end{bmatrix}\begin{bmatrix}
[NV^-](t)\\
[NV^0](t)
\end{bmatrix}.
\label{eqn:equation1}
\end{equation}
The transient model for the [NV$^-$] is then solved as:
\begin{equation}
    [NV^-](t)=\frac{\gamma_r}{\gamma_{tot}}+(\rho_0-\frac{\gamma_r}{\gamma_{tot}})e^{-\gamma_{tot}t}.
    \label{eqn:equation2}
\end{equation}
Here, $\gamma_i$ and $\gamma_r$ are the overall ionization and recombination rates respectively under multicolor excitation (Fig. \ref{fig:figure1}(f)), $\gamma_{tot}$ is the sum of $\gamma_i$ and $\gamma_r$, and $\rho_0$ is the [NV$^-$] at t = 0. The full expressions for $\gamma_i$ and $\gamma_r$ can be found in the Supplementary Material \cite{noauthor_see_nodate}. Using this model, we study the initialization fidelity as a function of the illumination powers and times (Fig. \ref{fig:figure2}(a, b)). We observe three general features: (i) The initialization rate depends on illumination power, and is more sensitive to NIR power (n). (ii) Higher fidelities can be achieved with lower green illumination power (g) and longer initialization time (t$_{ON}$). (iii) The fidelity has a non-monotonic dependence on the NIR illumination, and is maximized at a particular power -- the onset of the three-photon ionization process involving green and NIR lasers ($\gamma_{-,s}$ in Fig. \ref{fig:figure1}(f)) reduces the NV$^-$ initialization fidelity.\\
\indent 
Next, we analyze the experimental data and solve for the unknown rates ($\gamma_r$/$\gamma_{tot}$, $\gamma_{tot}$) in the model from fitting the model to the population data (dashed lines in Fig. \ref{fig:figure2}(a, b)). To verify the solved rates, we obtain a new dataset and draw the model for the new sets of powers that were not used during the fitting (Fig. \ref{fig:figure2}(c)). We see a good correspondence between the data and the model (see Supplementary Material \cite{noauthor_see_nodate} for more details on the rates solving process). We also note that we find our extracted rates to be consistently higher than those reported by Hopper et al. \cite{hopper2016near}, which could arise from differences in optical alignment and aberrations. We note that we do not observe linear ionization with power arising from surface defects, as has been observed in the past \cite{dhomkar2018charge}, likely because of the high quality surface terminations we utilize \cite{peace}. In general, the higher rates observed here overall are responsible for the faster NV$^-$ initialization time we report in this work.\\
\indent 
We then use the proposed model to map [NV$^-$] as a function of laser power and initialization time (Fig. \ref{fig:figure3}(a)-(d)). We see that at short t$_{ON}$ = 5 to 10 $\mu$s, [NV$^-$] $\geq$ 90\% can be achieved at high n, which is comparable to the sequence length for conventional NV center sensing protocols \cite{Hall2009Sensing,hall2016detection}. We combine the 2D plots into a single map to visualize the tradeoffs better in the parameter space (Fig. \ref{fig:figure3}(e)), where the experimental conditions for the data in Fig. \ref{fig:figure2}(c) are indicated by the black markers.

\begin{figure}
\centering
\includegraphics{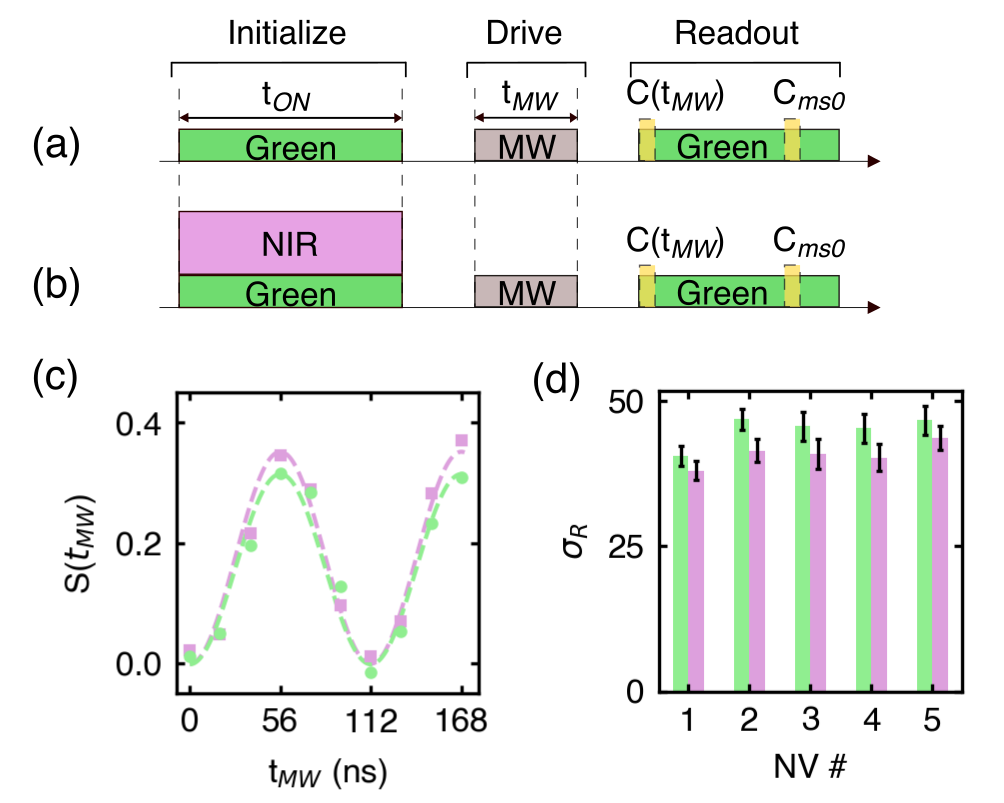}
\caption{\textbf{Characterization of spin readout.} (a), (b) Pulse sequence probing spin polarization after green or multicolor initialization (g = 100 $\mu$W, n = 2.9 mW, t$_{ON}$ = 100 $\mu$s) respectively. Microwave excitation (gray) resonant with the m$_s$ = 0 $\rightarrow$ -1 transition is applied during t$_{MW}$. C(t$_{MW}
$) and C$_{ms0}$ are the signal and reference photon counts during the 300 ns integration windows (yellow). (c) Spin contrast, S, as a function of t$_{MW}$ showing characteristic Rabi oscillation after green initialization (green circles) with contrast 32$\pm$2$\%$ and multicolor initialization (purple squares) with contrast 36$\pm$2$\%$. (d) Bar plot of the readout noise $\sigma_R$ for 5 NV centers for green (green bars) and multicolor (purple bars) initialization for each NV center.} 
\label{fig:figure4}
\end{figure}

\subsection*{C. Spin polarization after multicolor initialization}
For the multicolor initialization protocol to be useful in quantum applications, it must be compatible with the optical polarization of the NV$^-$ spin. To probe this, we perform a Rabi experiment using the pulse sequences shown in Fig. \ref{fig:figure4}(a, b). We choose conditions typical for conventional single-NV spin magnetometry (g = 100 $\mu$W, t$_{ON}$ = 100 $\mu$s, and n = 2.9 mW), so that [NV$^-$] increases from 80\% to 90\% after multicolor initialization. The spin contrast, S, is measured after a microwave (MW) pulse resonant with the m$_s$ = 0 $\rightarrow$ -1 transition:
\begin{equation}
    S(t_{MW})=\frac{C_{ms0}-C(t_{MW})}{C_{ms0}}.
    \label{eqn:equation3}
\end{equation}
Here, C(t$_{MW}$) and C$_{ms0}$ are the averaged signal and reference photon counts, and t$_{MW}$ is the MW pulse duration. We observe a 12.5$\%$ increase in the maximum contrast after the multicolor initialization, as shown in Fig. \ref{fig:figure4}(c), which is attributed to a relative increase in NV$^-$ photoluminescence over NV$^0$ photoluminescence (see Supplementary Material \cite{noauthor_see_nodate} for more analysis).
To further account for both the contrast and the photoluminescence change, we measure  the readout noise, $\sigma_R$ \cite{shields2015efficient}:
 \begin{equation}
    \sigma_R=\sqrt{1+2\frac{\alpha_0+\alpha_1}{(\alpha_0 - \alpha_1)^2}}.
    \label{eqn:equation4}
\end{equation}
Here, $\alpha_0$ = C$_{ms0}$/N and $\alpha_1$ = C(t$_\pi$)/N are the average number of photons collected in a single shot for the m$_s$ = 0 and m$_s$ = -1 states, respectively, N is the number of repetitions, and t$_\pi$ is the MW $\pi$-pulse duration. Multicolor initialization decreases the readout noise by 10\% on average, as shown in Fig. \ref{fig:figure4}(d), due to the 10\% increase in charge fidelity (see Supplementary Material \cite{noauthor_see_nodate} for more data and analysis on time-sensitivity \cite{hopper2018amplified}).

\begin{figure}
\centering
\includegraphics{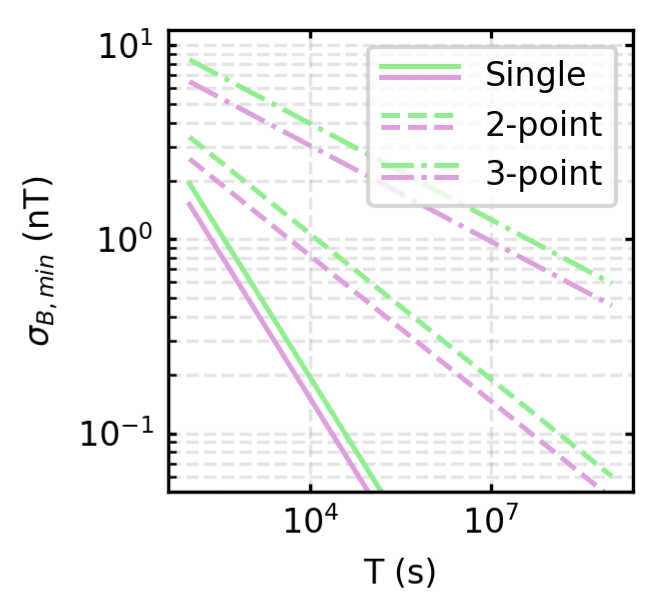}
\caption{\textbf{Multi-point correlations.} Minimum detectable magnetic field amplitude, $\sigma_{B,min}$ (Eqs. \ref{eqn:equation5} - \ref{eqn:equation7}) calculated as a function of total measurement time, T, for green (green lines) and multicolor (purple lines) initialization, with [NV$^-$] = 0.70 and 0.90 respectively, t$_{ON}$ = 5 $\mu$s and 50 $\mu$s respectively, t$_r$ = 2 ms, $\sigma_R$ = 4.8 and 3.7 respectively. Spin-to-charge conversion readout is considered for single NV center (solid lines), 2-point correlations (dashed lines), and 3-point correlations (dash-dot lines).} 
\label{fig:figure5}
\end{figure}

\subsection*{D. Sensitivity enhancement for conventional and covariance noise magnetometry}
The reduced readout noise has a greater impact on experiments measuring correlations among multiple NV centers simultaneously, such as in covariance magnetometry \cite{rovny2022nanoscale}. For a single NV center noise magnetometry experiment, the minimal detectable magnetic field amplitude is given by \cite{degen2017quantum,taylor2008high}:
\begin{equation}
    \sigma_{B,min}= \frac{\sigma_R}{4\gamma_e}\sqrt{\frac{t_m+t_r+t_{ON}}{T_2^2T}}.
    \label{eqn:equation5}
\end{equation}
Here, $\gamma_e$ is the gyromagnetic ratio of an electron, t$_m$ is the measurement time, t$_r$ is the readout time, t$_{ON}$ is the initialization time, T$_2$ is the decoherence time, and T = N(t$_m$+t$_r$+t$_{ON}$) is the overall experimental time. In this expression, the minimum detectable field scales as $\sigma_RT^{-1/2}$. However, measuring correlated magnetic noise sensed by two NV centers simultaneously under the same conditions is characterized by a stronger dependence on $\sigma_R$ \cite{rovny2022nanoscale}. Following the protocol shown in \textit{Rovny et al.}, in the long $T$ limit, we find the minimum detectable magnetic field amplitude for two-point (2p) and three-point (3p) correlations as:
\begin{align}
    \sigma_{B,min,2p}^2&= \frac{\pi\sigma_R^2{\rm e}^{2t_m/T_2}\cdot {\rm Hz}}{2\gamma_e^2t_m}\sqrt{\frac{t_m+t_r+t_{ON}}{T}}.
    \label{eqn:equation6} \\
    \sigma_{B,min,3p}^2&= \frac{\pi\sigma_R^2{\rm e}^{2t_m/T_2}\cdot {\rm Hz}}{2\gamma_e^2t_m}\left(\frac{t_m+t_r+t_{ON}}{T}\right)^{1/3}.
    \label{eqn:equation7}
\end{align}
We note that $\sqrt{\sigma_{B,min,2p}^2}\propto\sigma_RT^{-1/4}$ for two point correlations and $\sqrt{\sigma_{B,min,3p}^2}\propto\sigma_RT^{-1/6}$ for three-point correlations, evident in the slopes of the different lines in Fig. \ref{fig:figure5}, which shows the calculated minimal detectable magnetic field amplitude plotted as a function of T for the green ([NV$^-$] = 70\%) and multicolor initialization ([NV$^-$] = 90\%) (see Supplementary Material \cite{noauthor_see_nodate} for more details about readout noise in spin-to-charge readout and full expressions of minimum detectable magnetic field amplitudes). These differences in scaling result in different total experiment times to measure a field of a given amplitude. A 5-fold decrease in overall time is expected when using the multicolor initialization in a 3-point covariance magnetometry experiment.

Recent advances in camera technology allow multiplexing of tens of NV centers simultaneously, providing access to a multitude of many-point correlators \cite{cheng2024massively,cambria2024scalable}. The proposed optimized multicolor initialization protocol is compatible with the conventional wide-field optical microscope: 300 $\mu$s of 70 mW green and 800 mW NIR illumination of a 10$\times$10 $\mu$m$^2$ area results in [NV$^-$] = 95\%. The power demand can be further reduced by using spatial light modulators \cite{cheng2024massively} (1 mW green and 40 mW NIR for 50 NV centers) instead of wide-field illumination.

\section*{III. Conclusion and Outlook}
We have demonstrated a near-unity (95\%) charge state initialization of shallow NV centers under simultaneous 520 nm green and 905 nm NIR illumination with lower time and power overhead than previously reported for bulk NV centers \cite{hopper2016near}. This multicolor initialization allows NV$^-$ charge state initialization within 300 $\mu$s, and requires sub-mW NIR and a few $\mu$W green laser powers while preserving optical spin polarization. The initialization time can be as short as 10 $\mu$s for 90\% charge fidelity with marginally higher laser powers. These parameters are fully compatible with single-NV magnetometry and can be scaled to wide-field multiplexed excitation schemes.

The multicolor initialization scheme reduces SPAM errors, which is particularly advantageous for covariance noise magnetometry. The reduced SPAM error is critical for measuring multi-point correlators, opening the door to measuring new quantities in condensed matter physics and materials \cite{rovny2024new}. This scheme will also be useful for other applications that suffer from SPAM errors, such as sensing of correlated electric fields \cite{delord2024correlated,ji2024correlated} and NV-based quantum registers \cite{bradley2019ten,abobeih2022fault}.

\section*{Acknowledgements}
This work was primarily supported by the U.S. Department of Energy, Office of Science, and the Princeton Plasma Physics Laboratory under Contract Nop. DE-AC02-09CH11466. Sensitivity calculations and optimization was supported in part by the Center for Molecular Quantum Transduction (CMQT), an Energy Frontier Research Center funded by the U.S. Department of Energy, Office of Science, Basic Energy Sciences under Contract No. DE-SC0021314, and by the Gordon and Betty Moore Foundation and grant GBMF12237, DOI 10.37807. C.A.M. is supported by the U.S. Department of Energy, Office of Science, Office of Basic Energy Sciences, under Award No. DE-SC0012704. J.R. acknowledges support from the Intelligence Community Postdoctoral Research Fellowship Program by the Oak Ridge Institute for Science and Education (ORISE) through an interagency agreement between the US Department of Energy and the Office of the Director of National Intelligence (ODNI).\\ The authors also acknowledge the use of the Imaging and Analysis Center (IAC) operated by the Princeton Materials Institute at Princeton University, which is supported in part by the Princeton Center for Complex Materials (PCCM), a National Science Foundation (NSF) Materials Research Science and Engineering Center (MRSEC; DMR-2011750).
\section*{Author Declarations}
\subsection*{Conflict of interest}
The authors have no conflicts to disclose.
\subsection*{Author contributions}
M.M., A.L, and N.P.d.L. conceptualized the project, and designed experiments. M.M. carried out the experiments, analyzed the data and developed the rate equations model. A.L. and J.R. calculated sensitivity for noise magnetometry, Z.Y. contributed to the setup building. C.A.M. gave suggestions on the experiments and the manuscript. All authors contributed to writing and editing the manuscript.
\section*{Data availability}
The data that support the findings of this study are available from the corresponding author upon reasonable request.

\bibliography{reference}

\end{document}